\documentclass[aps,pra,twocolumn,a4paper,amsmath,amssymb,%
superscriptaddress,floatfix,]{revtex4}

\usepackage{graphicx,color}
\usepackage{epstopdf}

\newcommand{\Op}[1]{\boldsymbol{\mathsf{\hat{#1}}}}

\def\half{ \frac{1}{2}}
\def\openone{\leavevmode\hbox{\small1\kern-3.3pt\normalsize1}}

\begin{document}

\title{A Chebychev propagator for inhomogeneous Schr\"odinger equations}
\date{\today}

\author{Mamadou Ndong}
\affiliation{Institut f\"ur Theoretische Physik,
  Freie Universit\"at Berlin,
  Arnimallee 14, 14195 Berlin, Germany}
\author{Hillel Tal-Ezer}
\affiliation{Academic College of Tel-Aviv Yaffo, Israel}
\author{Ronnie Kosloff}
\affiliation{Department of Physical Chemistry and
  The Fritz Haber Research Center,
  The Hebrew University, Jerusalem 91904, Israel}
\author{Christiane P. Koch}
\email{ckoch@physik.fu-berlin.de}
\affiliation{Institut f\"ur Theoretische Physik,
  Freie Universit\"at Berlin,
  Arnimallee 14, 14195 Berlin, Germany}

\begin{abstract}
  A propagation scheme for time-dependent inhomogeneous
  Schr\"odinger equations is presented. Such equations occur in 
  time dependent optimal control theory and  
  in reactive scattering. A formal solution based on a polynomial 
  expansion of the inhomogeneous term is derived. It is subjected
  to an approximation in terms of Chebychev polynomials. Different
  variants for the inhomogeneous propagator are demonstrated and
  applied to two examples from optimal control theory.
  Convergence behavior and numerical efficiency are analyzed. 
\end{abstract}

\maketitle


\section{Introduction}
\label{sec:intro}

Inhomogeneous time-dependent Schr\"odinger equations,
\begin{equation}
  \label{eq:inhtdSE}
  i\hbar\frac{\partial}{\partial t} |\psi(t)\rangle = \Op{H}
  |\psi(t)\rangle +  \Op{G}(t)|\varphi(t)\rangle\,,
\end{equation}
arise in many formal solutions of quantum dynamics. In particular they 
have been employed in a time dependent
treatment of reactive scattering \cite{NeuhauserCPC91,NeuhauserCPL92}
and in optimal control theory using time-dependent targets
\cite{KaiserJCP04,SerbanPRA05,KaiserCP06,WerschnikJPB07}
or state-dependent constraints \cite{JoseMyPRA08}.
In reactive scattering, the inhomogeneity results from the application
of a projection operator \cite{NeuhauserJCP89}. This projector divides
the Hilbert space of the reactive system into subspaces
corresponding, respectively, to the  reactants and to the products.
A reduced description for only the products can be derived where the
time-dependent Schr\"odinger equation contains an inhomogeneity,
i.e. a source term that
corresponds to the creation of the products \cite{NeuhauserCPL92}.

In optimal control theory (OCT), the inhomogeneity may be caused by a
projection operator as well. For example, a partitioning of the 
Hilbert space is implemented by a projection operator in order to
suppress population in a forbidden subspace  \cite{JoseMyPRA08}. 
This leads to a formulation of OCT with a state-dependent
constraint containing the projection operator. As a result the backward 
propagation of the OCT equations includes an
inhomogeneity in the Schr\"odinger equation. This term 
corresponds to the suppression of probability
amplitude in the forbidden subspace. 

Generally,  an
inhomogeneous Schr\"odinger  equation arises in OCT if
a time-dependent target or a state-dependent constraint are utilized.
In the common versions of OCT, see 
e.g. Refs. \cite{ZhuJCP98,JosePRA03}, the target is not explicitly
time-dependent, it depends only on some final time $T$. The constraints
enforce the Schr\"odinger equation and a minimization of the field
energy. But for explicitly time-dependent targets
\cite{KaiserJCP04,SerbanPRA05,KaiserCP06,WerschnikJPB07} or for a state-dependent
constraint \cite{JoseMyPRA08,OhtsukiJCP04} the
 optimization functional contains a contribution of the form
\[
\lambda \int_0^T g\left[\psi(t),\psi^*(t)\right] dt\,,
\]
where the state $|\psi\rangle$ of the system enters at each time $t$.

For the solution of the standard homogeneous time-dependent
Schr\"odinger equation, a number of numerical propagation schemes exist
\cite{RonnieReview88,RonnieReview94}. The Chebychev propagator
\cite{RonnieCheby84} offers the advantage of a numerically exact
solution. The accuracy of the calculation is then determined by the
machine precision of the computer and the error is uniformly
distributed. The propagator is based on approximating the formal solution of the
homogeneous time-dependent Schr\"odinger equation,
\begin{equation}
  \label{eq:formalsolhomo}
  |\psi(t+dt)\rangle =  e^{-\frac{i}{\hbar}\Op{H}dt} |\psi(t)\rangle \, ,
\end{equation}
by a series of Chebychev polynomials.
Time-dependent \textit{inhomogeneous} Schr\"odinger equations have been
solved to date with  split-propagator schemes \cite{SerbanPRA05} or
via a full diagonalization of the Hamiltonian
\cite{JoseMyPRA08}.
While  the latter method is numerically expensive
and quickly becomes unfeasible with increasing system size, the first
is of only limited accuracy.

Here, we derive a formal solution of the time-dependent inhomogeneous
Schr\"odinger equation and we adapt it to the Chebychev propagation scheme. 
We apply this new propagator to the optimal control with a
state-dependent constraint and with a time-dependent target.
The paper is organized as follows: Section~\ref{sec:formalsol}
presents the formal solution of Eq.~(\ref{eq:inhtdSE}). 
Propagation schemes for the formal solution are derived in
Section~\ref{sec:cheby}. The Chebychev propagation scheme 
is applied to OCT with a state-dependent
constraint where the system is forced to remain in a subspace of the
total Hilbert space in Section~\ref{sec:appl1}.
In this case the operator $\Op{G}$ in
Eq.~(\ref{eq:inhtdSE}) is independent of time, $\Op{G}(t)=\Op{G}$.
In Section~\ref{sec:appl2}, a second application, OCT with a
time-dependent target, is studied
keeping the full time-dependence, $\Op{G}(t)$. In both applications,
the convergence of the new Chebychev propagator is analyzed in detail.
Section~\ref{sec:concl} concludes.


\section{Formal solution}
\label{sec:formalsol}

The inhomogeneous Schr\"odinger equation, Eq.~(\ref{eq:inhtdSE}), is
treated as an ordinary differential equation. It can
be rewritten (setting $\hbar=1$)
\begin{equation}
  i\frac{\partial}{\partial t} |\psi(t)\rangle =
  \Op{H}  |\psi(t)\rangle +  |\Phi(t)\rangle\,,
  \label{eq:inhSE}
\end{equation}
where $|\Phi(t)\rangle = \Op{G}(t)|\varphi(t)\rangle$.
Eq. (\ref{eq:inhSE}) is  solved subject to the boundary conditions
\begin{equation}
  \label{eq:ini}
  |\psi(0)\rangle =  |\psi_0\rangle\,, \quad\quad
  |\Phi(0)\rangle =  |\Phi_0\rangle\,.
\end{equation}
$|\Phi(t)\rangle$ is known globally in the propagation time interval,
$[0,T]$, for example by a numerical representation on $N_t$ sampling
points. $|\Phi(t)\rangle$ is assumed to be analytic, 
so that it can be interpolated to any arbitrary point within $[0,T]$. 
The representation of $|\Phi(t)\rangle$ on $N_t$ sampling points
corresponds to an expansion 
in $N_t$ basis functions. Choosing equidistant sampling points yields
a Fourier representation. A high-order (usually $N_t \gg 1$)
polynomial expansion is obtained when the sampling points are chosen
as roots of polynomials, implying non-equidistant time steps. 
The optimal representation treating correctly the boundaries is
obtained by choosing Chebychev polynomials.  

The basic idea consists in devising a short-time integration scheme
for the  interval $[0,t]$ (or, more generally $[t_n,t_{n+1}]$
with $n, n+1 \le N_t$) by taking the following polynomial expansion
of the inhomogeneous term,
\begin{equation}
|\Phi(t)\rangle \approx
\sum_{j=0}^{m-1} P_j(\bar{\tau}) |\bar{\Phi}_j\rangle \equiv
\sum_{j=0}^{m-1} \frac{t^j}{j!} |\Phi^{(j)} \rangle\,.
\label{eq:Taylor}
\end{equation}
Here, $P_j$ denotes the Chebychev polynomial of order $j$ with
expansion coefficient $|\bar{\Phi}_j\rangle$, and
$\bar{\tau}$ is a rescaled time,  $\bar{\tau} = \frac{2\tau}{ t}-1$,
for $\tau \in [0,t]$.
Note that the sum on the right-hand side of Eq.~(\ref{eq:Taylor}) is
only of the form of a Taylor expansion for later manipulations but the
approach itself is not based on a Taylor expansion.
The Chebychev expansion coefficients
$|\bar{\Phi}_j\rangle$ can be obtained e.g. by 
a cosine transformation (cf. Ref.~\cite{RoiPRA00} and
Section~\ref{sec:inhomcheby} below).
Once the coefficients $|\bar{\Phi}_j\rangle$ 
are known they are used to generate the coefficients $|\Phi^{(j)} \rangle$
in the second sum of Eq. (\ref{eq:Taylor}), cf. Appendix \ref{app:trafo}.
This procedure has the advantage of employing a uniform approximation
of $|\Phi^{(j)} \rangle$ in the interval $[0,t]$. 
However, if the function $|\Phi(t)\rangle$ is not known
analytically, it needs to be interpolated at
sampling points $\tau \in [0,t]$ in order to calculate the expansion
coefficients $|\bar{\Phi}_j\rangle$. As a simpler alternative,
the function $|\Phi^{(j)} \rangle$ is expanded in a Taylor series in the
time interval $[0,t]$. Then the coefficients
$|\Phi^{(j)}\rangle$ become the $j$th derivative of $|\Phi(t)\rangle$,
at the beginning of the interval. 
At this point the properties of the global $N_t$ interpolation
function of $|\Phi(t)\rangle$ are used to calculate $|\Phi^{(j)}\rangle$
as numerical derivatives at the beginning of each interval.

Based on Eq. (\ref{eq:Taylor}), the solution of Eq.~(\ref{eq:inhSE})
can be written as 
\begin{equation}
  |\psi(t)\rangle_{(m)} = \sum_{j=0}^{m-1} \frac{t^j}{j!} |\lambda^{(j)} \rangle +
  \Op{F}_m |\lambda^{(m)}\rangle\,.
\label{eq:formalsol}
\end{equation}
In Eq.~(\ref{eq:formalsol}), the subscript $m$ denotes the order of
the solution.
The $|\lambda^{(j)}\rangle$ are obtained iteratively,
\begin{eqnarray}
  \label{eq:lambdas}
  |\lambda^{(0)} \rangle &=& |\psi_0 \rangle\, , \\
  |\lambda^{(j)}  \rangle &=& -i \Op{H}|\lambda^{(j-1)} \rangle +
  |\Phi^{(j-1)}  \rangle, \nonumber\\ && \quad\quad 1\leq j \leq m \,,\nonumber
\end{eqnarray}
and $\Op{F}_m$ is a function of $ \Op{H}$ given by
\begin{equation}
  \label{eq:Fm}
  \Op{F}_m = f_m(\Op{H}) = (-i\Op{H})^{-m} \left (e^{-i\Op{H} t} -
    \sum_{j=0}^{m-1} \frac{(-i \Op{H} t)^j}{j!} \right )\,.
\end{equation}
Equations~(\ref{eq:formalsol})-(\ref{eq:Fm}) represent the formal
solution to the inhomogeneous Schr\"odinger equation.

In order to verify that Eq.~(\ref{eq:formalsol}) is indeed a
solution to Eq.~(\ref{eq:inhSE}), let's take the derivative of
Eq.~(\ref{eq:formalsol}),
\begin{widetext}
\begin{eqnarray*}
  \frac{\partial}{\partial t} |\psi(t)\rangle_{(m)} &=&
  \sum_{j=0}^{m-1} j\frac{t^{j-1}}{j!} |\lambda^{(j)}  \rangle +
  (-i\Op{H})^{-m} \left ((-i\Op{H})e^{-i\Op{H} t} -
    \sum_{j=1}^{m-1} j (-i\Op{H}) \frac{(-i \Op{H} t)^{j-1}}{j(j-1)!} \right )
  |\lambda^{(m)}  \rangle   \\
  &=&  \sum_{j=0}^{m-2} \frac{t^{j}}{j!} |\lambda^{(j+1)}  \rangle
  + (-i\Op{H}) (-i\Op{H})^{-m}  \left (e^{-i\Op{H} t} -
    \sum_{j=0}^{m-2}  \frac{(-i \Op{H}t)^{j}}{j!}\right )
  |\lambda^{(m)}  \rangle \,.
\end{eqnarray*}
Inserting Eq.~(\ref{eq:lambdas}) in the first term and resumming with
an upper limit $m-1$, one obtains
\begin{eqnarray*}
  \frac{\partial}{\partial t} |\psi(t)\rangle_{(m)}
  &=&  \sum_{j=0}^{m-1} \frac{t^{j}}{j!} \left(-i
  \Op{H}|\lambda^{(j)}  \rangle  + |\Phi^{(j)}  \rangle\right)
  - \frac{t^{m-1}}{(m-1)!} \left(
    -i \Op{H} |\lambda^{(m-1)}  \rangle + |\Phi^{(m-1)}  \rangle\right)   \\
  && \quad\quad\quad -i\Op{H}  (-i\Op{H})^{-m}  \left (e^{-i\Op{H} t} -
    \sum_{j=0}^{m-1}  \frac{(-i \Op{H}t)^{j}}{j!} +
    \frac{(-i \Op{H}t)^{m-1}}{(m-1)!} \right )|\lambda^{(m)}  \rangle\,. \\
 \end{eqnarray*}
\end{widetext}
Recognizing $\Op{F}_m$, cf. Eq.~(\ref{eq:Fm}), as part of the last
term and inserting $|\lambda^{(j)}  \rangle$, cf. Eq.~(\ref{eq:lambdas}), in
the second term, which then cancels with the second summand of the last
term, this can be rewritten
 \begin{eqnarray*}
  \frac{\partial}{\partial t} |\psi(t)\rangle_{(m)}
  &=& -i \Op{H} \left(\sum_{j=0}^{m-1} \frac{t^{j}}{j!} |\lambda^{(j)} 
  \rangle +\Op{F}_m |\lambda^{(m)}  \rangle\right) \\
&&+  \sum_{j=0}^{m-1} \frac{t^{j}}{j!}|\Phi^{(j)}  \rangle \,.
\end{eqnarray*}
The expression in parenthesis corresponds to
$|\psi(t)\rangle$, cf. Eq.~(\ref{eq:formalsol}). Finally, replacing
the second term  by
$|\Phi(t)\rangle$, cf. Eq.~(\ref{eq:Taylor}), the inhomogeneous
Schr\"odinger equation, Eq.~(\ref{eq:inhSE}), is obtained. Therefore
Eq.~(\ref{eq:formalsol}) presents indeed a solution to Eq.~(\ref{eq:inhSE}).


\section{Propagation schemes}
\label{sec:cheby}

The formal solution, Eq.~(\ref{eq:formalsol}), is subjected to a
spectral approximation,
\begin{equation}
  \label{eq:specapprox}
  |\psi(t)\rangle_{(m)N} = \sum_{j=0}^{m-1} \frac{t^j}{j!} |\lambda^{(j)}  \rangle +
  P_{N(m)}(\Op{H}) |\lambda^{(m)} \rangle\,,
\end{equation}
where $P_N$ is a polynomial of order $N$ approximating $\Op{F}_m=f_m(\Op{H})$. 
For example, to first order, $m =1$, the formal solution is given by
\begin{equation}
  |\psi(t)\rangle_{(1)} = e^{-i\Op{H} t} |\psi_0\rangle+
  (-i\Op{H})^{-1} (e^{-i\Op{H} t} - \openone)
  |\Phi_0\rangle \,.
\label{eq:firstorder}
\end{equation}
In principle one might seek a spectral approximation for each of the
terms. However, it is numerically more efficient to rewrite the first order solution,
\begin{equation}
  \label{eq:newfirstorder}
  |\psi(t)\rangle_{(1)} = |\psi_0\rangle+
f_1(\Op{H})\left(-i\Op{H}|\psi_0\rangle + |\Phi_0\rangle \right) \,,
\end{equation}
with $ f_1(\Op{H})= (-i\Op{H})^{-1} (e^{-i\Op{H} t} - \openone)$, 
such that only a single Chebychev expansion plus one extra application of the
Hamiltonian are required.
Similarly, to second and third order the formal solutions can be written
\begin{eqnarray}
  \label{eq:secondorder}
  |\psi(t)\rangle_{(2)}&=& |\psi_0\rangle +
t |\lambda^{(1)}\rangle  +  f_2(\Op{H}) |\lambda^{(2)}\rangle  \,,\\
   \label{eq:thirdorder}
 |\psi(t)\rangle_{(3)}&=& |\psi_0\rangle+ 
 t |\lambda^{(1)}\rangle  + 
  \frac{t^2}{2} |\lambda^{(2)}\rangle \nonumber \\
  &&+ f_3(\Op{H})  |\lambda^{(3)}\rangle
\end{eqnarray}
with
$f_2(\Op{H})$ and $f_3(\Op{H})$ given by Eq.~(\ref{eq:Fm}) and
$|\lambda^{(j)}\rangle$ by Eq.~(\ref{eq:lambdas}).
The strategy is then to seek a polynomial approximation for the
functions $f_j(\Op{H})$. For $f_0(\Op{H}) = e^{-i\Op{H} t} $, this
corresponds to the standard Chebychev 
propagator for the homogeneous Schr\"odinger equation \cite{RonnieCheby84}. It will be
briefly reviewed for clarity, followed by a discussion of the polynomial
approximation of the new functions.

\subsection{General idea of the Chebychev propagator}

The Chebychev propagator is based on treating the formal solution as a
function of an operator which is applied to some state vector,
\[
|\phi\rangle = f(\Op{H})|\psi\rangle\,,
\]
and to approximate this function by an expansion in Chebychev
polynomials $P_n$,
\[
f(\Op{H})|\psi\rangle  = \sum_n a_n P_n(\Op{H}) |\psi\rangle \,.
\]
Since the Chebychev polynomials are defined within the range $[-1,1]$,
the Hamiltonian has to be renormalized,
\[
\Op{H}_{norm} = 2 \frac{\Op{H}-E_{min}\openone}{\Delta E} -\openone\,,
\]
where $E_{min}$ denotes the smallest eigenvalue of $\Op{H}$ and
$\Delta E=E_{max}-E_{min}$ the spectral range of $\Op{H}$.
The wavefunction propagated from time zero to $t$ is then obtained by
\[
  |\psi(t)\rangle \approxeq
  e^{-i(\half\Delta E+E_{min}) t}  \sum_{n=0}^N a_n
  P_n(-i\Op{H}_{norm})|\psi_0\rangle  \,,
\]
where the phase factor in front of the sum is due to the renormalization.

The algorithm to estimate $|\phi\rangle = f(\Op{H})|\psi\rangle$
proceeds as  follows:
\begin{enumerate}
 \item Calculate the expansion coefficients $a_n$,
   \begin{equation}
     \label{eq:an}
     a_n = \frac{2-\delta_n}{\pi} \int_{-1}^1
     \frac{f(x)P_n(x)}{\sqrt{1-x^2}} dx  \,.
   \end{equation}
   For the function $f(x) = f_0(x) = e^{-ixt}$, the integrals can be solved
   analytically resulting in the Bessel functions
   \cite{RonnieCheby84}.
 \item Calculate $P_n(-i\Op{H}_{norm})|\psi_0\rangle$
   using the recursion relation of the Chebychev polynomials,
   \begin{eqnarray}
     \label{eq:chebyrec}
     |\phi_0\rangle &=& |\psi_0\rangle \,,\nonumber \\
     |\phi_1\rangle &=& -i \Op{H}_{norm} |\psi_0\rangle \,,\\
     |\phi_{n}\rangle &=& -2i \Op{H}_{norm} |\phi_{n-1}\rangle +
     |\phi_{n-2}\rangle\,. \nonumber
   \end{eqnarray}
 \item Accumulate the result, taking into account the phase factor due
   to renormalization,
   \[
   |\psi(t)\rangle = e^{-i(\half\Delta E+E_{min}) t} \sum_{n=0}^N a_n |\phi_n\rangle\,.
   \]
\end{enumerate}
Task 1 has to be performed only once, while 2 and 3 are repeated
for each propagation step. The number $N$ of Chebychev polynomials is
chosen such that the coefficient $a_{N+1}$ becomes smaller than some
specified error $\varepsilon$. Since the coefficients can be
determined analytically, $\varepsilon$ may correspond to the machine
precision of the computer.

\subsection{Chebychev propagator for inhomogeneous equations}
\label{sec:inhomcheby}

According to Eqs.~(\ref{eq:firstorder}-\ref{eq:thirdorder})
a single  Chebychev expansion plus a few applications of the Hamiltonian
are required to solve the inhomogeneous Schr\"odinger equation. In
particular, the application of a function $f_m( \Op{H})$ to a state
vector $|\lambda^{(m)}\rangle$ has to be considered,
\begin{equation}
  \label{eq:phi}
    |\phi\rangle = f_m(\Op{H}) |\lambda^{(m)} \rangle\,,
\end{equation}
where $|\lambda^{(m)}\rangle$ is obtained recursively by
Eq.~(\ref{eq:lambdas}). 
The Chebychev propagation scheme now consists in calculating
$|\lambda^{(m)} \rangle$ and performing tasks 1--3 
for $f_m(\Op{H}) |\lambda^{(m)}\rangle$ instead of $f_0(\Op{H})
|\lambda^{(m)}\rangle$. 
There are two 
differences with respect to the standard Chebychev propagator,
i.e. between the approximation of $f_0(\Op{H})$ and the approximation
of $f_{m>0}(\Op{H})$. 

(i) For $m>0$, the integrals required to calculate the expansion
coefficients $a_n$, Eq.~(\ref{eq:an}), cannot be solved
analytically anymore. They can, however, be obtained numerically with
sufficient efficiency and accuracy \cite{RoiPRA00}.
The approach of Ref. \cite{RoiPRA00} is repeated here for completeness:
Applying a Gaussian quadrature to the Chebychev polynomials, 
the integral in Eq.~(\ref{eq:an}) is rewritten,
\begin{equation}
  \label{eq:an_general}
  a_n = \frac{2-\delta_{n0}}{N}\sum_{k=0}^{N-1} f_m(x_k) P_n(x_k) \,,
\end{equation}  
where the sampling points $x_k$ correspond to the $N$ roots of the
Polynomial $P_N$, 
\begin{equation}
  \label{eq:xk}
  x_k = \cos \left(\frac{\pi(k+\frac{1}{2})}{N}  \right)\,, k=0,\,\ldots\, N-1 \,.
\end{equation}  
Since the Chebychev polynomials can be
expressed in terms of cosines, $P_n(x) = \cos(n\theta)$ with $\theta
=\arccos(x)$,  Eq.~(\ref{eq:an_general}) corresponds to a 
cosine transformation, 
\begin{equation}
 \label{eq:an_cos}
  a_n = \frac{2-\delta_{n0}}{N}\sum_{k=0}^{N-1} f_m(\theta_k) \cos(n\theta_k) \,.
\end{equation}  
The expansion coefficients are thus most easily evaluated
by Fast Cosine Transformations.

Special care is, however, required for small values of the argument of
$f_{m}(\Op{H})$ since $f_{m}(\Op{H})$ involves division by the
$m$th power of $\Op{H}$. It is recommended to employ the definition of
$f_{m}(\Op{H})$, Eq.~(\ref{eq:Fm}), only if the argument is larger than some
small value $\epsilon$ and to use a Taylor expansion of $f_{m}(\Op{H})$,
\begin{equation}
  \label{eq:FmTaylor}
  f_{m>0}(\Op{H}) = t^m \sum_{j=0}^\infty \frac{(-i\Op{H}t)^j}{(j+m)!}
\end{equation}
for arguments smaller than  $\epsilon$.

(ii) As explained in Section~\ref{sec:formalsol}, a Taylor expansion of
$|\Phi(t)\rangle$ for each integration interval $[0,t]$ is most easily
used if the inhomogeneous term is represented numerically at sampling
points.  
Recursive calculation of  $|\lambda^{(m)}\rangle$ then requires
numerical evaluation of the time derivatives of the 'inhomogeneous state
vector', $|\Phi(t)\rangle$. Second order $m=2$ requires the first derivative 
which can be obtained with sufficient accuracy by Fast Fourier
Transformation (FFT) and multiplication in frequency domain. However,
an error is introduced due to finite values of  $|\Phi(t)\rangle$ at
the boundaries of the grid, i.e. $t=0$ and $t=T$. This error is increased and
propagated throughout the time interval in the calculation of  higher
order derivatives  by FFT and multiplication in frequency
domain.

Higher order schemes require therefore a different
method for the evaluation of the derivatives. A suitable choice represents
$|\Phi(t)\rangle$ by an expansion into Chebychev polynomials. The
derivatives are then calculated recursively based on the analytical
properties of the Chebychev polynomials \cite{Dunn96}. Note that this
implies a non-equidistant time grid since the roots of the Chebychev
polynomials need to be taken as sampling points,
\begin{eqnarray*}
t'_n = \cos\left(\frac{n\pi}{N_t-1}\right)
\,,\,
t_n &=& \frac{T}{2}\left(1+\cos\left(\frac{n\pi}{N_t-1}\right)\right)\,,  \\
n &=& 0, \ldots, N_t-1\,,
\end{eqnarray*}
where the time interval $t \in [0,T]$ is scaled to $t'\in [-1,1]$
and $N_t$ denotes the number of sampling points. A
non-equidistant time grid requires calculation and storage of the
Chebychev expansion coefficients of the propagator for each $\Delta
t_n$. This additional effort is well paid off 
since a higher order scheme allows for larger time steps, i.e. 
smaller $N_t$.


\subsection{Summary of the  Algorithm}
\label{sec:summaryalgorithm}

The algorithm to solve the inhomogeneous time-dependent Schr\"odinger
equation is summarized in order to outline the flow chart of the numerical
implementation: 
\begin{enumerate}
\item Set the time grid $\{t_n\}$ and determine the inhomogeneous term
  $|\Phi(t_n)\rangle$.  
\item Calculate the Chebychev expansion coefficients of
  $\Op{F}_m=f_m(\Op{H})$, cf. Eq.~(\ref{eq:Fm}). This needs to be done
  for each time step size that occurs in a time grid with non-equidistant
  steps or only once if an equidistant time grid is employed.
\item Determine all $|\Phi^{(j)}  \rangle$ required in
  Eqs.~(\ref{eq:formalsol}) and (\ref{eq:lambdas}). This is done
  differently whether the uniform approximation or the Taylor
  expansion over the short-time interval $[0,t]$ is employed.
  
  Uniform approximation (Chebychev expansion):
  \begin{enumerate}\renewcommand{\theenumii}{\roman{enumii}}
    \item Determine Chebychev sampling points $\{\tau_k\}$
      within each short-time
      interval  $[0,t]$ (or, generally, $[t_n,t_{n+1}]$),
      cf. Eq.~(\ref{eq:xk}), and determine the value of 
      $|\Phi(\tau_k)\rangle$ analytically or by numerical
      interpolation. 
    \item Calculate the Chebychev expansion coefficients
      $|\bar{\Phi}_j\rangle$ of $|\Phi(\tau_k)\rangle$ 
      by cosine transformation  
      analogously to Eq.~(\ref{eq:an_cos}) with $f_m$ replaced by
      $|\Phi\rangle$. 
    \item Derive $|\Phi^{(j)}\rangle$ for all $\{t_n\}$
      from $|\bar{\Phi}_j\rangle$
      using the transformation between a Chebychev and a Taylor
      expansion given in Appendix~\ref{app:trafo}, i.e. employing the 
      recursive relations given in Eqs.~(\ref{eq:coeffB0}) and
      (\ref{eq:coeffBi}). 
  \end{enumerate}
     
  Taylor expansion: \\
  Calculate  $|\Phi^{(j)}\rangle$ as time derivatives of
  $|\Phi(t_n)\rangle$ either  
  by FFT for equidistant time grid or by numerical differentiation
  based on the analytical 
  properties of  Chebychev polynomials \cite{Dunn96} for non
  equidistant time steps. 
\item Perform the time propagation, i.e. for each time step from 0 to
  $t$ (or, generally, $t_n$ to $t_{n+1}$),
  \begin{enumerate}\renewcommand{\theenumii}{\roman{enumii}}
   \item determine  $|\lambda^{(j)}  \rangle$ according to
     Eq.~(\ref{eq:lambdas}), 
   \item obtain $\Op{F}_m |\lambda^{(j)}\rangle$ by Chebychev
     recursion, cf. Eq.~(\ref{eq:chebyrec}), 
   \item evaluate $|\Psi(t)\rangle$ (or, generally, $|\Psi(t_{n+1})\rangle$)
     according to  Eq.~(\ref{eq:formalsol}).
   \end{enumerate}       
\end{enumerate}
In order to avoid storage of the Chebychev expansion coefficients,
step 2 in the case of a non-equidistant time grid and step 3 in the
case of the uniform approximation may also be performed during the
time propagation in step 4. 

The algorithm to solve the inhomogeneous time-dependent Schr\"odinger
equation requires adjustment of two parameters -- the order of the
algorithm, $m$, and the propagation time step, i.e. the size of the
short-time interval. The latter can also be specified in terms of the
overall number of time steps, $N_t$, which is particularly useful for
non-equidistant time steps. The two parameters are linked to each
other: Employing a higher order $m$ allows for taking larger time
steps, or, equivalently, reducing $N_t$. A detailed discussion how the
two parameters should be chosen is presented for the applications in
Sections~\ref{sec:appl1} and \ref{sec:appl2} below.

\subsection{Approximation of the formal solution by a Taylor expansion}
\label{sec:taylor}

A simplified version of the algorithm is devised by approximating 
the formal solution explicitly by a Taylor expansion. This yields a
standard Chebychev propagator plus additional terms involving
derivatives of $|\Phi(t)\rangle$. An existing Chebychev propagation
code then needs little modification to be adapted to an inhomogeneous
Schr\"odinger equation. The accuracy of the simplified algorithm is,
however, limited by the Taylor expansion.

As shown in Appendix~\ref{app:approx},
the formal solution, Eq.~(\ref{eq:formalsol}), can be written as
\begin{equation}
  |\psi(t)\rangle_{(m)} = e^{-i\Op{H} t} |\psi_0 \rangle +
  \sum_{j=0}^{m-1}  \Op{F}_{j+1}  |\Phi^{(j)}  \rangle\,,
\label{eq:formalsolbis}
\end{equation}
with
\[
\Op{F}_{j+1}  = (-i\Op{H})^{-(j+1)} \left (e^{-i\Op{H} t} -
                \sum_{k=0}^{j} \frac{(-i \Op{H} t)^k}{k!} \right ).
\]
Taking the Taylor expansion of the exponential in
$\Op{F}_{j+1}$ up to order $j+1$,
\[
e^{-i\Op{H} t} =  \sum_{k=0}^{j+1} \frac{(-i \Op{H} t)^k}{k!}
=  \sum_{k=0}^{j} \frac{(-i \Op{H} t)^k}{k!} +  \frac{(-i \Op{H} t)^{j+1}}{(j+1)!}  \,,
\]
one obtains an approximation of Eq.~(\ref{eq:formalsolbis}),
\begin{equation}
 |\psi(t)\rangle_{(m)} \approx e^{-i\Op{H} t} |\psi_0 \rangle +
  \sum_{j=0}^{m-1}  \frac{t^{j+1}}{(j+1)!}  |\Phi^{(j)}  \rangle\,.
 \label{eq:approxformalsol}
\end{equation}
To all orders, only the standard Chebychev  
propagation scheme plus calculation and storage of 
the derivatives of the inhomogeneous term, $|\Phi^{(j)}\rangle$, are required.
For example, the approximate solution to second order  reads
\begin{equation}
  |\psi(t)\rangle_{(2)} \approx f_0(\Op{H})  |\psi_0\rangle+
  t|\Phi^{(0)}\rangle + \frac{t^2}{2}|\Phi^{(1)} \rangle\,.
\label{eq:secondorderbis}
\end{equation}

Unlike the propagator described in Sections~\ref{sec:inhomcheby} and
\ref{sec:summaryalgorithm}, this 
scheme is explicitly based on a Taylor expansion. It is therefore
expected to be valid only for small time steps. In that case, a low
order scheme (e.g. $m=2$) is sufficient. The necessary derivatives can
then be calculated by FFT and multiplication in frequency
domain. The validity of this approximation is discussed in detail in
Section~\ref{subsec:test}.


\section{Application I: Control with a state-dependent constraint}
\label{sec:appl1}

In our first example, the operator occuring in the inhomogeneous term
is not time-dependent, $\Op{G}(t)=\Op{G}$.

\subsection{Model}
A simplified model of the vibrations in a Rb$_2$ molecule is
considered taking into account three electronic states
\cite{JoseMyPRA08}. The corresponding Hamiltonian reads
\begin{eqnarray}
  \label{eq:H}
\Op{H}&\,=\,& \sum_{i=1}^{3}\Op{H}_i\otimes |e_i\rangle\langle e_i| +
\Op{\mu}\,\epsilon(t) \cdot \,\bigg(|e_1\rangle\langle e_2|+ \\
&& |e_2\rangle\langle e_1|+
|e_2\rangle\langle e_3|+|e_3\rangle\langle
e_2|\bigg)\,, \nonumber
\end{eqnarray}
where $\Op{H}_i$ denotes the vibrational Hamiltonians,
$\Op{H}_i=\Op{T}+V_i(\Op{R})$, $\Op{\mu}$ the transition dipole
operator, assumed to be independent of $\Op{R}$, and $\epsilon(t)$ the
electric field.
The electronic states are associated to
$X^1\Sigma_g^+(5s+5s)$, $^1\Sigma_u^+(5s+5p)$
and $^1\Pi_g(5s+4d)$, with the
potentials found in Ref.~\cite{ParkJMS01}.

OCT is tested  for the objective of population transfer from the vibrational
ground state of the electronic ground surface to a particular vibrationally
excited state via Raman transitions between the ground and the second
electronic surface. For strong laser fields $\epsilon(t)$, such as
those found by OCT algorithms,
population at intermediate times will be excited not only to the
second but also to the third electronic surface. This is particularly
the case for the electronic states of our example, where transition
frequencies and Franck-Condon factors for transitions between
$|e_1\rangle$ and $|e_2\rangle$ are very similar to those for
transitions between $|e_2\rangle$ and $|e_3\rangle$.
Assuming that the
third electronic state corresponds to a loss channel, e.g. an
intermediate state in resonantly enhanced multi-photon ionization or
an autoionizing state, population transfer into this state should be
avoided at all times. This can be communicated to the OCT algorithm
by formulating a state-dependent constraint which maximizes the
projection onto the allowed subspace \cite{JoseMyPRA08},
i.e. onto electronic states 1 and 2. The complete functional for
optimization is then given by \cite{JoseMyPRA08}
\begin{equation}
  \label{eq:functional_j}
  J[\varphi,\varphi^*,\epsilon] =
  J_0[\varphi_T,\varphi_T^*]\,+\,J_a[\epsilon] \,+\,J_b[\varphi,\varphi^*]
\end{equation}
with
\begin{eqnarray}
J_0[\varphi_T,\varphi_T^*]&=&\lambda_0\,
\langle\varphi(T)|\Op{D}|\varphi(T)\rangle \,, \\
J_a[\epsilon] &=&\int_0^T \lambda_a(t)\,[\epsilon(t)-\epsilon_r(t)]^2
\, dt \,,\\
J_b[\varphi,\varphi^\dagger]&=&\int_0^T
\lambda_b\langle\varphi(t)|\Op{P}_\mathrm{allow}|\varphi(t)\rangle \,dt \,.
\end{eqnarray}
The operator $\Op{D}$ in $J_0$ is given by the projector onto the target level 
in the electronic ground state. The state-dependent constraint
contains the projector onto the allowed subspace, $\Op{P}_\mathrm{allow}$.
In Ref. \cite{JoseMyPRA08},
propagation via full diagonalization of the Hamiltonian was employed and
only 11 vibrational levels in each electronic state were considered.
Representing the vibrational Hamiltonians on a Fourier grid
\cite{RonnieReview88} and
utilizing the inhomogeneous Chebychev propagator, the full potentials
can be taken into account ($N_\mathrm{grid}=128$).

\subsection{Test of the propagator}
\label{subsec:test} 
In order to test the new Chebychev propagator, our
numerical results are compared to those obtained by the
symmetrical method for the Hamiltonian comprising of 11 levels in each
electronic state \cite{JoseMyPRA08}.
The details of the symmetrical method are reviewed in
Appendix~\ref{app:symm_method}. 
The inhomogeneous Schr\"odinger equation reads \cite{JoseMyPRA08}
\begin{equation}\label{eq:chistate}
\frac{d}{dt}|\psi(t)\rangle\,=\,
-\frac{i}{\hbar}\,\Op{H}[\epsilon(t)]\,
|\psi(t)\rangle\,+\,\lambda_b\Op{P}_\mathrm{allow}|\varphi(t)\rangle\,,
\end{equation}
with the ``initial'' condition
\begin{equation}\label{eq:chi_T}
|\psi(t=T)\rangle\,=\,-\lambda_0\,\Op{D}\,|\varphi(T)\rangle\,.
\end{equation}
In OCT, $|\psi(t)\rangle$ is the backward propagated wavefunction
which is coupled to $|\varphi(t)\rangle$ obtained by forward propagation,
\begin{equation}\label{eq:schrodinger}
\frac{d}{dt}|\varphi(t)\rangle\,=\,-\frac{i}{\hbar}\Op{H}[\epsilon(t)]
|\varphi(t)\rangle
\end{equation}
with $|\varphi(t=0)\rangle\,=\,|\varphi_0\rangle$.
The convergence with respect to
the time step $\Delta t$ and the order $m$ 
of the Chebychev method will be discussed in Section
~\ref{subsec:conv1}.

We start by propagating the initial state, $v=0$ of the 
electronic ground state, forward with a Gaussian pulse given by Eq.~(33) of Ref.
\cite{JoseMyPRA08}. The inhomogeneous Schr\"odinger equation is then
propagated backward with the same field. All parameters 
are taken to be equal to those of Ref. \cite{JoseMyPRA08}.
\begin{figure}[tb]
  \centering
  \includegraphics[width = 0.9\linewidth]{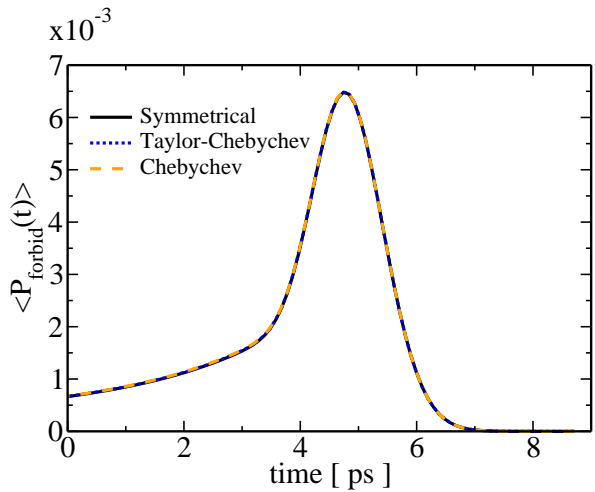}
  \caption{(color online)
    Normalized  expectation value of the projector onto the forbidden
    subspace  for a single backward propagation with a Gaussian
    pulse.  Comparison between  the symmetrical
    scheme of Ref. \cite{JoseMyPRA08} with  $\Delta t =
    1\,$a.u. (black solid line), 
    the Chebychev propagator in first order with $\Delta t =
    1\,$a.u. (orange dashed line) and  
    the Chebychev propagator in first order based on the Taylor
    approximation, Eq.(\ref{eq:approxformalsol}), with $\Delta t =
    0.1\,$a.u. (blue dotted line).
  } 
  \label{fig:cheb_symm}
\end{figure}
Figure \ref{fig:cheb_symm} shows the  expectation value of
the projector onto the  forbidden subspace, $\langle\Op{P}_{forbid}
\rangle(t)$, for  normalized $|\psi(t)\rangle$.
Results for the symmetrical method and the first order 
Chebychev propagators based on Eq.~(\ref{eq:formalsol}) and 
on Eq.~(\ref{eq:approxformalsol}) (Taylor) are compared. 
A good agreement between the Chebychev propagators and the symmetrical
scheme is found.  
The time step needed to obtain converged results with the
approximate solution is, however, smaller by a factor of ten.
This is not surprising since the approximate solution is based on a
Taylor expansion which requires a very small $\Delta t$.

Our main motivation for the development of the modified Chebychev
propagator lies in its application in OCT with a state-dependent
constraint or a time-dependent target. Therefore the 
new propagator is also tested in an iterative solution of the control
equations. The control problem is that of Ref.~\cite{JoseMyPRA08}.
However, in the present work the full
vibrational Hamiltonians are employed. 
\begin{figure}[tb]
  \centering
  \includegraphics[width = 0.9\linewidth]{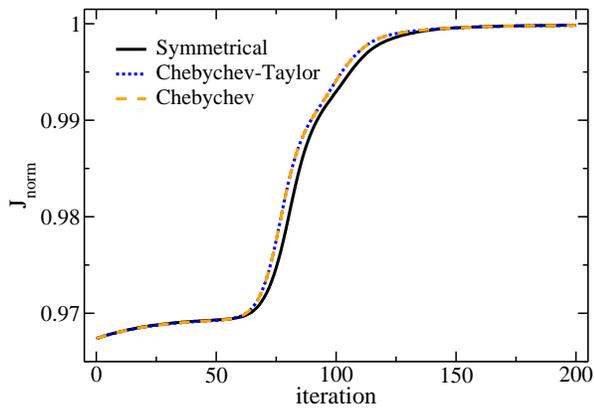}
  \caption{(color online) $J_\mathrm{norm}$ as a function
    of the number of iterations.  Solution obtained by the symmetrical scheme
    of Ref. \cite{JoseMyPRA08} with a time step $\Delta t = 1\,$a.u. (black solid
    line),  the  Chebychev propagator for first
    order with a time step  $\Delta t = 1\,$a.u. (orange dashed line) 
    and the second order approximate 
    solution of Eq.(\ref{eq:approxformalsol}) with a time step $\Delta t = 0.1\,$a.u.
    (blue dotted line).
  } 
  \label{fig:OCT_result}
\end{figure}

Convergence of the OCT algorithm is measured by the
normalized functional, $J_\mathrm{norm}= J/(\lambda_0+\lambda_bT)$,
approaching one as the control equations are iteratively solved,
cf. Fig.~\ref{fig:OCT_result}.
The symmetrical scheme ($\Delta t= 1\,$a.u., black solid line) is
compared to the Chebychev propagator in first order (orange dashed
line) and to the Chebychev propagator based on the Taylor 
approximation, Eq.~(\ref{eq:approxformalsol}),
(blue dotted line) in second order.
An overall good agreement is found.
The difference between the results at intermediate iterations is 
attributed to the different models, the full vibrational
Hamiltonian in the case of the Chebychev propagators and the
Hamiltonian consisting of 33 levels in the case of the symmetrical method. 

Application of the Chebychev propagator based on the Taylor
approximation, Eq.~(\ref{eq:approxformalsol}), in first order failed:
The inhomogeneous Schr\"odinger equation is solved with insufficient
accuracy and the property of monotonic convergence of the Krotov method
\cite{JosePRA03} is lost during the iterative solution of the control
equations. 
This is easily rationalized in terms of the very fast oscillations
occuring in the optimized field. They require either a very
small time step or a higher order in the approximation of the exponential,
$e^{-i\Op{H} \Delta t}$, by a Taylor expansion.
The complexity of the pulse and hence the time variation of
the inhomogeneous term determine the required order of the Taylor
approximation.

\subsection{Convergence behavior}
\label{subsec:conv1}

The convergence and the efficiency 
of the inhomogeneous Chebychev propagator are analyzed with respect
to the number of time steps $N_t$ and the order $m$ of the
solution. The main numerical effort is required for the application of
the Hamiltonian and the calculation of the derivatives. For a given
propagation time $T$ one would like to identify optimum values of
$N_t$ and $m$ that yield a mininum computation time for a specified
accuracy. In general, decreasing the number of time steps $N_t$
or, respectively, increasing $\Delta t$
will require a larger number of Chebychev polynomials in the expansion
of the function $f_m(\Op{H})$, but also 
a higher order $m$ of the solution. The recursive
calculation of the Chebychev polynomials in the expansion of $f_m(\Op{H})$
implies continued application of the
Hamiltonian. Moreover, higher order solutions require additional
applications of the Hamiltonian,
cf. Eqs.~(\ref{eq:firstorder}-\ref{eq:thirdorder}), and determination 
of derivatives of the inhomogeneous term
up to degree $m-1$. 

\begin{figure}[tb]
  \centering
  \includegraphics[width = 0.9\linewidth]{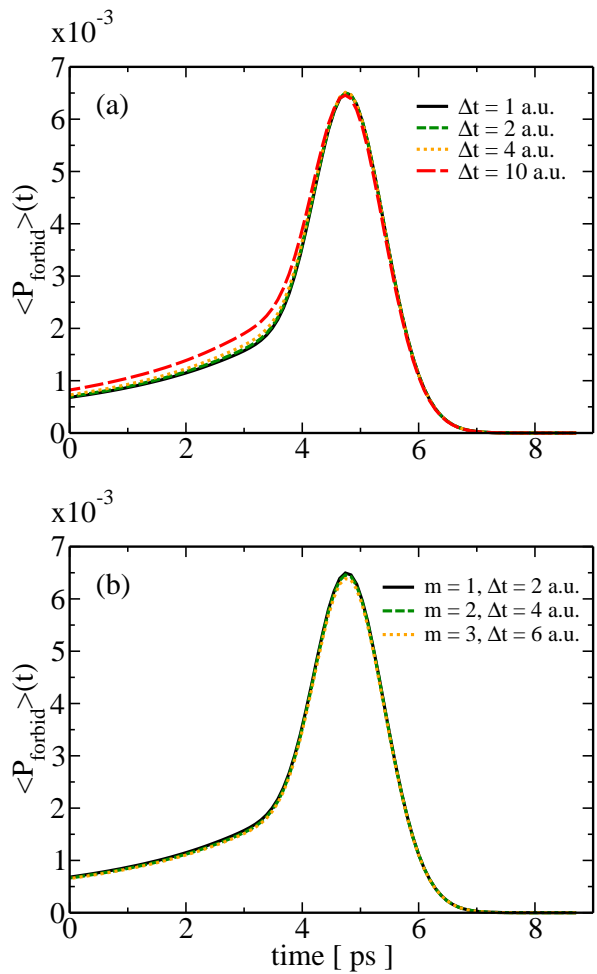}
  \caption{(color online)
    Normalized expectation value for the projector onto the forbidden
    subspace demonstrating the convergence of the inhomogeneous
    Chebychev propagator: (a) for first order solutions ($m=1$),
    convergence is lost by increasing $\Delta t$, 
    (b) converged results for the largest possible time step at a
    given order $m$. 
  }
  \label{fig:convergence}
\end{figure}

We consider first equidistant time steps and evaluate the time derivatives
by FFT. Figure \ref{fig:convergence} traces the population in the forbidden
subspace to illustrate the convergence behavior for dynamics under a
Gaussian pulse.   
In first order, cf. Fig.~\ref{fig:convergence}a,
the dynamics are found to be converged for time steps
$\Delta t\leq 2\,$a.u. At larger time steps, deviations are observed,
in particular at short times (occuring late for backward
propagation). 
Dynamics with the largest possible time step at each order
are shown in Fig.~\ref{fig:convergence}b.
\begin{table}[tb]
  \centering
  \begin{tabular}{|c|c|c|c|c|c|}\hline
    order $m$ & $\Delta t$ & $N_t$ & $N_\mathrm{Cheby}$ & applications of $\Op{H}  $ & CPU time \\ \hline
    1  & $2\,$a.u.  &180.000 & 6     & $1.260.000$  &            $12\,$s \\
    2  & $4\,$a.u.  &90.000 & 8    & $900.000 $               & $11\,$s\\ 
    3  &$6\,$a.u.   &60.000 & 10     & $780.000  $              & $9\,$s\\
    \hline
  \end{tabular}
  \caption{CPU time required for one backward propagation
    to obtain converged solutions in first,
    second and third order for a total propagation time of $8\,$ps
    with equidistant time steps $\Delta t$.
    Also listed are the number of time steps, $N_t$, the
    number of times the Hamiltonian is
    applied and the number of terms in the Chebychev expansion, $N_\mathrm{Cheby}$.
  }
  \label{tab:cpu}
\end{table}

In order to decide whether it is numerically more efficient to keep a
low order demanding a small time step or to employ a higher order allowing
for a larger time step, Table~\ref{tab:cpu} compares 
the CPU time required to obtain converged solutions in first, 
second and third order. The number of applications of $\Op{H}$
includes both those occuring in the Chebychev recursion and those due
to the additional terms in
Eqs.~(\ref{eq:firstorder}-\ref{eq:thirdorder}). 
For example, for third order and $\Delta t = 6\,$a.u.,
10 Chebychev polynomials are sufficient to approximate $f_3(\Op{H})$ 
Eq.~(\ref{eq:thirdorder}). Each time step then implies
thirteen applications of $\Op{H}$,
ten
for the Chebychev recursion plus three for the additional terms
(one for $|\lambda^{(1)}\rangle$ and two for $|\lambda^{(2)}\rangle$),
cf. Eqs.~(\ref{eq:thirdorder}) and (\ref{eq:lambdas}).
As can be seen in Table~\ref{tab:cpu}, 
in terms of CPU time, it is more efficient to employ a higher order solution.
In the context of OCT calculations, in addition to saving computation time, a
higher order propagator also allows for saving memory since
the backward propagated wavefunction needs to be stored for each time step.
An inherit limit to increasing the time step is, however, posed by the
time-dependence of the Hamiltonian. Expressing the formal solution of
the homogeneous time-dependent Schr\"odinger equation by the
exponential, $e^{-i\Op{H}\Delta t}$, assumes $\Op{H}$ to be constant
within the time interval $\Delta t$. In our example the upper limit at
third order, $\Delta t=6\,$a.u., is due to the breakdown of this
assumption for the forward propagated wavefunction, $|\varphi(t)\rangle$,
entering the inhomogeneous term,
$\lambda_b\Op{P}_\mathrm{allow}|\varphi(t)\rangle$.

\begin{figure}[tb]
  \centering
  \includegraphics[width = 0.9\linewidth]{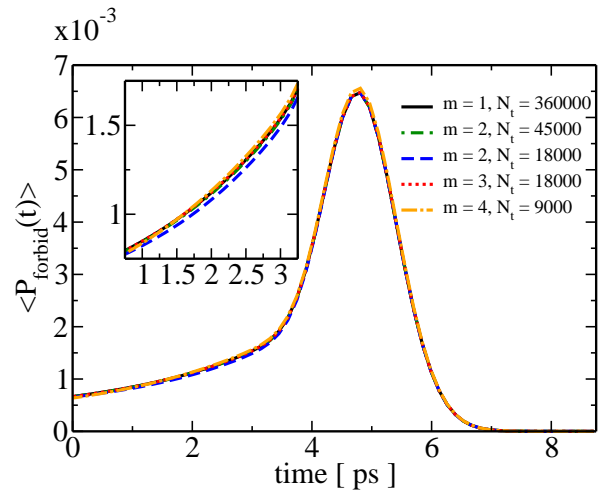}
  \caption{(color online) Normalized expectation value of the
    projector onto the forbidden subspace for a Hamiltonian with slow
    time-dependence (RWA) employing a non-equidistant time grid and
    calculating time derivatives in terms of Chebychev expansions for
    different orders $m$. Results are shown for the
    smallest possible number of time steps, 
    $N_t$, at given order $m$ except for $m=2$, $N_t=18000$ (blue
    dashed line) which illustrates a non-converged case. 
  }
  \label{fig:convergence:nonequi}
\end{figure}

In order to demonstrate that a higher-order solution for the
inhomogeneous Schr\"odinger equation allows indeed for a large time
step, we modify our example by invoking the rotating-wave
approximation (RWA). This eliminates highly oscillatory terms from the
field, $\epsilon(t)$, keeping only the time-dependence of the
envelope which is several orders of magnitude slower. However, when
increasing the time step and the order, numerical determination of the
derivatives by FFT and multiplication in frequency domain breaks down, 
cf. Section~\ref{sec:inhomcheby}. Accurate numerical calculation of
the time derivatives of the inhomogeneous term,
$|\Phi(t)\rangle=\lambda_b\Op{P}_\mathrm{allow}|\varphi(t)\rangle$, is afforded by
expanding $|\Phi(t)\rangle$ in Chebychev polynomials. The sampling
points of the time grid then need to be chosen as the roots of the
Chebychev polynomials, leading to non-equidistant time
steps (dividing the time interval into equidistant time steps
corresponds to a Fourier representation).
Figure~\ref{fig:convergence:nonequi} and Table~\ref{tab:cpu:nonequi}
illustrate the convergence behavior for a Hamiltonian with slow
time-dependence. 
\begin{table}[tb]
  \centering
  \begin{tabular}{|c|c|c|c|c|}\hline
    order $m$ & $N_t$ &  $\Delta t_\mathrm{max}$ &CPU time \\ \hline
    2         & $45.000$ & $12.6\,$a.u. &    $170\,$s \\
    3         & $18.000$ & $31.4\,$a.u. & $69\,$s\\ 
    4         & $9.000$  & $62.8\,$a.u. & $35\,$s\\
    \hline
  \end{tabular}
  \caption{CPU time required to obtain converged solutions in 
    second, third and fourth order for a total propagation time of $8\,$ps
    with non-equidistant time steps. Also listed are the smallest
    possible number of sampling points for the time grid, $N_t$, and the
    corresponding maximum time step, $\Delta t_\mathrm{max}$.
  }
  \label{tab:cpu:nonequi}
\end{table}
All results in Fig.~\ref{fig:convergence:nonequi} are shown for the
smallest possible number of time steps except the blue dashed line
($m=2$, $N_t=18.000$) which illustrates a non-converged case, cf. the
deviation from the converged results at short times. 
The number of time steps can be significantly reduced by employing
higher-order schemes. The evaluation of the derivatives by Chebychev
expansion is, however, more costly, leading to overall larger
computation times than in Table~\ref{tab:cpu}. It is obvious from
Table~\ref{tab:cpu:nonequi} that this variant of the inhomogeneous
Chebychev propagator will unfold its full power for a time-independent
Hamiltonian that occurs e.g. in reactive scattering calculations where
a very large time step together with a high-order scheme will be
numerically most efficient. The fact that the permissible time step
can be increased by employing a higher-order solution illustrates
that inhomogeneous Chebychev propagator is based on a global representation. 

\section{Application II: Control with a time-dependent target}
\label{sec:appl2}

In our second application, the operator occuring in the inhomogeneous term,
$\Op{G}(t)$,  is explicitly time-dependent.

\subsection{Model}
In principle it should be possible to prescribe by a laser pulse an
arbitrary pathway that the quantum system should follow. To this end, OCT with a
time-dependent target has to be employed \cite{SerbanPRA05,WerschnikJPB07}.
In the total functional, Eq.~(\ref{eq:functional_j}),
the final-time term then disappears, $J_0[\varphi_T,\varphi_T^*]=0$, and
the state- and time-dependent term becomes \cite{JoseMyPRA08}
\begin{equation}
  \label{eq:Jb2}
  J_b[\varphi,\varphi^\dagger] = \int_0^T
  \lambda_b\langle\varphi(t)|\Op{G}(t)|\varphi(t)\rangle \,dt \,.
\end{equation}
Maximization of $J_b$ corresponds to $\lambda_b \ge 0$ and fulfills
the conditions for monotonic convergence \cite{JoseMyPRA08}.

\begin{figure}[tb]
  \centering
  \includegraphics[width=0.9\linewidth]{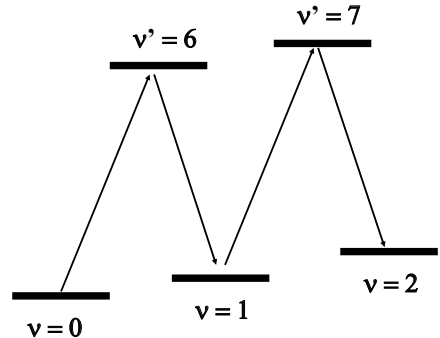}
  \caption{Prescribed 'trajectory' for a time-dependent target: climbing up the
    ladder of a double $\Lambda$-system.
    }
  \label{fig:scheme}
\end{figure}

A simple model comprising of five of the 33 levels of
Section~\ref{sec:appl1} are taken to mimic a double  $\Lambda$-system,
cf. Fig.~\ref{fig:scheme}.
Initially all population is assumed to be in $v=0$, and at the final
time the population in $v=2$ is to be maximal. Additionally,
the time interval $[0,T]$ is divided into subintervals where the
population of the intermediate levels $v'=6$, $v=1$, and $v'=7$, is
maximized, i.e. we prescribe a 'trajectory' where the ladder of the
double $\Lambda$-system is sequentially climbed up. While this
represents a simple toy model, it serves the purpose of illustrating the
case where the operator of the inhomogeneous term of the
Schr\"odinger equation, $\Op{G}(t)$, is explicitly time-dependent.
The inhomogeneous equation for backward propagation reads,
\begin{equation}\label{eq:chistatetimetarg}
\frac{d}{dt}|\psi(t)\rangle\,=\,
-\frac{i}{\hbar}\,\Op{H}[\epsilon(t)]\,
|\psi(t)\rangle\,+\,\lambda_b\Op{G}(t)|\varphi(t)\rangle\,,
\end{equation}
with the ``initial'' condition
\begin{equation}\label{eq:chi_Ttimetarg}
|\psi(t=T)\rangle\,=\, 0\,.
\end{equation}

Dividing the time interval $[0,T]$ into four subintervals, 
 $0<T_1<T_2<T_3<T$, 
the target is defined as the projector onto $v'=6$ in 
$[0,T_1]$, onto  $v=1$ in  $[T_1,T_2]$, onto
 $v'=7$ in   $[T_2,T_3]$ and onto  $v=2$ in the subinterval $[T_3,T]$,
\begin{eqnarray}
\label{eq:deftimetarget}
\Op{G}(t) &=& |6\rangle\langle 6|\Theta(T_1-t)+\nonumber\,\\
&&
|1\rangle\langle 1|\Theta(t-T_1)\Theta(T_2-t) +\nonumber\,\\ 
&&
|7\rangle\langle 7|\Theta(t-T_2)\Theta(T_3-t) \nonumber\,\\
&&
|2\rangle\langle 2|\Theta(t-T_3)\Theta(T-t) 
\end{eqnarray}
with $\Theta(t)$ the Heaviside function. In order to avoid numerical
problems due to discontinuities, $\Theta(t)$ is approximated by
\begin{equation} 
\Gamma(t) = \frac{1}{1+e^{-kt}} \,,
\end{equation}
where the parameter $k$ determines the steepness with which the target
level changes. While for large values of $k$, the step function is
recovered, small values of $k$ imply overlap in time of two different
targets near the $T_i$, $i=1-3$. In the following $k$ is varied
between $k=10^{-4}\,$a.u. and $k=10^4\,$a.u.
The final time is set to $T = 5.4\,$ps.  The subintervals are taken to
be of the same length, 
 $T_1 = 1.35\,$ps,  $T_1 = 2T_1$ and  $T_3 = 3T_1$.

\subsection{Results}
\label{subsec:res}
\begin{figure}[tb]
  \centering
  \includegraphics[width=0.9\linewidth]{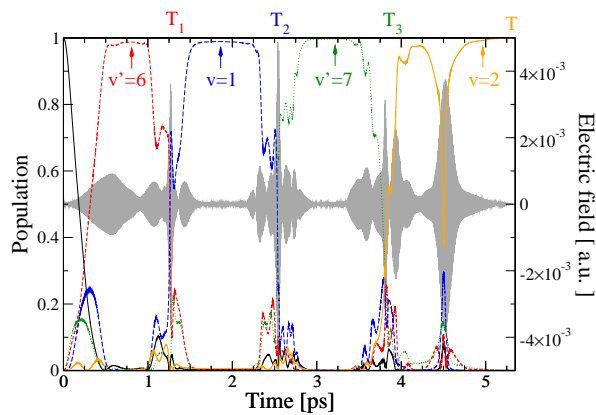}
  \caption{(color online) Time-dependent target:
    Time evolution of the level populations with an optimal field
    (shown in grey). 
    }
8  \label{fig:poptimetarget}
\end{figure}

\begin{figure}[tb]
  \centering
  \includegraphics[width=0.9\linewidth]{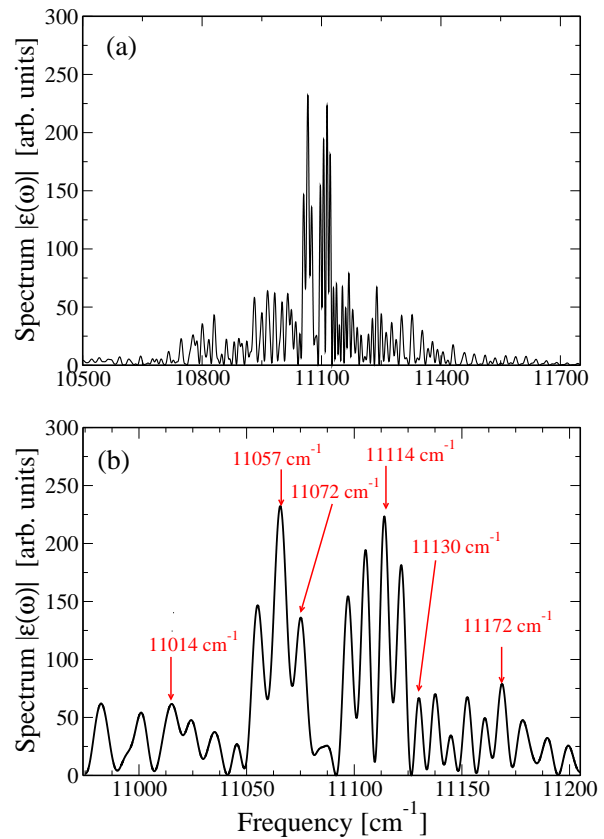}
  \caption{(color online) (a) Spectrum of the optimized field. (b) Detail
  of the spectrum ($\omega \in [10975\,\mathrm{cm}^{-1},11205\,\mathrm{cm}^{-1}]$). 
  The numbers and arrows indicate the six main transition frequencies
  of the model, cf. Table~\ref{tab:freqtransition}.}
  \label{fig:spectimetarget}
\end{figure}

The new propagator is  tested for the iterative solution of the control
equations with the time-dependent target. 
The guess field consists of a sequence of four $\pi$-pulses, one in
each time interval.  Figure~\ref{fig:poptimetarget}
shows the evolution of the level populations using the optimized
field for $k=10^4$. 
They follow by and large indeed the prescribed 'trajectory'.
Population of levels other than the target one and fast oscillations
in the populations are observed only when switching from one target to
the next. 
The spectrum of the optimized field is shown in Fig.~\ref{fig:spectimetarget}.
The transition frequencies of our model, listed  in
Table~\ref{tab:freqtransition} and indicated in
Fig.~\ref{fig:spectimetarget}b, are contained within the
spectrum.
Additional frequencies which do not correspond to the main transition
frequencies are observed. They are attributed to the complexity of the
optimal solution which may include beatings between levels, Stark
shifts etc.
\begin{table}[tb]
  \centering
  \begin{tabular}{|cc|c|c|c|}\hline
        &  & 0 & 1 & 2 \\ \hline
    6  && $ 11130\,$ & $ 11072\,$ & $ 11014\,$ \\
    7  && $ 11172\,$ & $ 11114\,$ & $ 11057\,$ \\ \hline
  \end{tabular}
  \caption{Transition frequencies in cm$^{-1}$ for the five levels
    employed as time-dependent target.}
  \label{tab:freqtransition}
\end{table}

\begin{figure}[tb]
  \centering
  \includegraphics[width=0.9\linewidth]{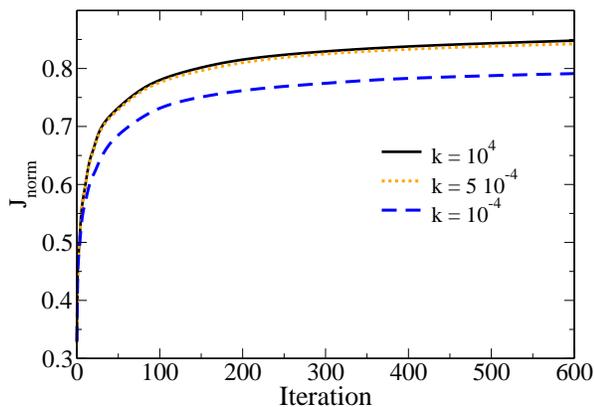}
   \caption{(color online)
     The renormalized functional $J_b$, cf. Eq.~(\ref{eq:Jb2}),
     as a function of
     the number of iterations for parameters $k$ corresponding to 
     different overlaps in time of two targets.
   }
   \label{fig:Jtimetarget}
\end{figure}

The improvement of the time-dependent target functional with the
number of iterations is demonstrated in Fig.~\ref{fig:Jtimetarget} for
different values of the steepness parameter $k$.
Monotonic convergence is observed.
However, the algorithm cannot reach 100\%. We attribute this to the
way the target is switched and overlap in time of different targets
is created around the $T_i$, $i=1-3$. 
For large $k$ the changes in the target functional are almost
instantaneous and cannot be followed by the dynamics, cf. the
oscillations in the level populations in
Fig.~\ref{fig:poptimetarget}. However, at the same time, the targets
do almost not overlap.
This yields the highest value of the target functional, about
85\%. 
For smaller values of $k$ the dynamics can follow more
smoothly. However, the overlap between different targets is increased,
i.e. contradictory objectives are asked at the same time. This
decreases the value of the target functional to about 79\%.

\subsection{Convergence behavior}
\label{subsec:conv2}
\begin{figure}[tb]
  \centering
  \includegraphics[width=0.9\linewidth]{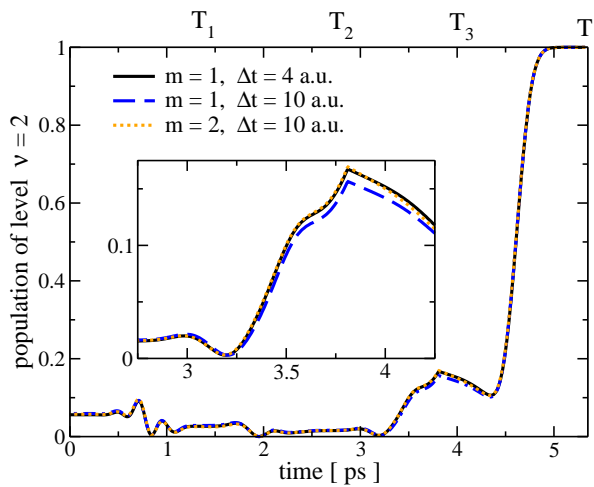}
  \caption{(color online) (a) Time-dependent population of the final target
level $\nu\,=\,2$: Converged results in first and second orders for the largest
possible time step (black solid and orange dotted lines).
Also shown is a non-converged result in first order (blue dashed line) - the deviations
from the converged solutions are evident in the insert. }
  \label{fig:pop_nu2firstsecond}
\end{figure}

The convergence of the inhomogeneous Chebychev propagator is again analyzed
with respect to the time step $\Delta t$ and to the order $m$. We
restrict ourselves here to the case of equidistant time steps and
calculation of the derivatives by FFT and multiplication in frequency
domain.
The convergence behavior is illustrated in
Fig.~\ref{fig:pop_nu2firstsecond} by 
the time evolution of the  final target level ($v=2$) population. 
Converged results  are obtained for  $\Delta t \le 4\,$a.u.
($\Delta t \le 10\,$a.u.) in first (second) order,
i.e. a larger time step than in Section~\ref{sec:appl1}
can be used. We attribute this to the much simpler model. 

Table~\ref{tabtimetarg:cpu} compares the CPU time required to obtain
converged solutions in first and second order. 
\begin{table}[tb]
  \centering
   \begin{tabular}{|c|c|c|c|c|}\hline
    order $m$ & time step $\Delta t$ & $N_\mathrm{Cheby}$ & applications of $\Op{H}  $ & CPU time \\ \hline
    1                   & $4\,$a.u.  & 8     & $495.000$  &            $0.72\,$s \\
    2                   & $10\,$a.u.  & 10    & $264.000 $               & $0.62\,$s\\ 
    \hline
   \end{tabular}
  \caption{CPU time required to obtain converged solutions in first
    and second order for a total propagation time of $5.4\,$ps}
  \label{tabtimetarg:cpu}
\end{table}
The same conclusion  is obtained as in Section~\ref{sec:appl1}, i.e it
is more  efficient to employ a higher order scheme.

Overall, no difference in the convergence behavior for time-dependent
and time-independent operators in the inhomogeneous term, $\Op{G}(t)$
is found. This can be rationalized as follows: the convergence is
determined by the fastest timescale of the dynamics, i.e. by the rapid
oscillations of the field. The time-dependence of the projection
operator introduces a time-dependence which is much slower and hence
does not affect convergence.

\section{Conclusions}
\label{sec:concl}

A formal solution to the time-dependent inhomogeneous Schr\"odinger
equation was derived based on an expansion of the
inhomogeneous term. Three levels of Chebychev approximations are
involved. 
\begin{itemize}
\item[(i)] The first one yields the Chebychev propagator where the
argument of the Chebychev polynomials is the Hamiltonian.
Truncating the expansion at the desired order $m$,
the formal solution is subjected to a spectral representation 
with Chebychev polynomials. A propagation scheme similar to the
standard Chebychev propagator for homogeneous Schr\"odinger equations
is then obtained: Instead of $e^{-i\Op{H}\Delta t}$ a function $f_m(\Op{H})$
is expanded in Chebychev polynomials. For the exponential function, the
expansion coefficients can be calculated analytically, for the
function $f_m$they need to be
obtained numerically. This is achieved by Fast Cosine
Transformations, utilizing the definition of Chebychev polynomials
in term of cosines. 
\item[(ii)] The second level expands the inhomogeneous state vector
$|\Phi(t)\rangle$ in Chebychev polynomials within each short-time
integration interval $[0,t]$. The argument of this expansion is the
rescaled time 
$\bar{t}$ covering the interval. This Chebychev approximation
is easily applied only if $|\Phi(t)\rangle$ is known analytically. If
$|\Phi(t)\rangle$ is determined numerically on sampling points
covering the global propagation time interval $[0,T]$, there are two
choices. $|\Phi(t)\rangle$  needs to be interpolated to sampling 
points within $[0,t]$. In a simpler alternative, the Chebychev 
expansion is replaced by a Taylor expansion based on numerical 
derivatives at the beginning of each time step. This has been 
done for the present applications.
\item[(iii)] The numerical calculation of the derivatives requires a third
level of Chebychev approximation where the argument is the time $t$
covering the global
propagation time interval $[0,T]$. This implies a non-equidistant time
grid where the derivatives are evaluated according to the procedure
described in Ref.~\cite{Dunn96}. This expansion overcomes the
numerical error introduced by non-zero boundary values of the
inhomogeneous state vector $|\Phi(t)\rangle$ at $t=0$ and $t=T$. An
alternative based on equidistant time steps employs  Fast Fourier Transforms and
multiplication in frequency domain. However, in that case the 
errors introduced at the boundary of the time
grid build up. Therefore this scheme is limited to low order where
only first or second derivatives are required. 
\end{itemize}

An even more approximate solution to the  time-dependent inhomogeneous Schr\"odinger
equation is obtained by rewriting
the formal solution explicitly in terms of a Taylor expansion. The
propagator then consists of the standard exponential term plus time
derivatives of the inhomogeneous terms. 
This approximation is numerically less efficient than the
propagator for the full formal solution. Moreover, it may become
instable in optimal control applications where the inhomogeneous
term often is highly oscillatory and the numerical evaluation of 
derivatives by FFT becomes difficult. The main advantage of this propagation
scheme lies in the fact that it requires very little modification of
existing standard Chebychev propagation codes. 

Both Chebychev propagation schemes were tested in two optimal control
applications. OCT with a state-dependent constraint
\cite{JoseMyPRA08}, e.g. maximizing 
population in an allowed subspace of the Hilbert space, yields a
time-independent operator in the inhomogeneous term while an
explicitly  time-dependent operator is obtained  in OCT
with a time-dependent target
\cite{KaiserJCP04,SerbanPRA05,KaiserCP06,WerschnikJPB07}. Convergence
of the propagation schemes was demonstrated for both applications. The
convergence behavior was studied in detail as a function of the order
of the solution and the required number of time steps for a given
overall propagation time. For applications with a fast
time-dependence such as OCT, a low order scheme with a small time step and
evaluation of the time derivatives by FFT was found to be the best
choice. For applications with a slow time-dependence or
time-independent Hamiltonians such as reactive scattering calculations 
where large time steps are permissible, 
a high-order scheme is numerically most efficient.
This reflects that the propagation scheme is based on a global
representation of the inhomogeneous term. It is this regime
where the new propagator can best unfold its power. 

The new Chebychev propagator provides a stable and accurate numerical
solution to the time-dependent inhomogeneous Schr\"odinger equation. 
It is most efficient for high order and large time steps. Ideally an
inherent time-dependence of the Hamiltonian should also be
incorporated into the Chebychev scheme. This is the subject of a further
study.


\begin{acknowledgments}
  We would like to thank Jos\'e Palao for stimulating discussions.
  Financial support from the  Deutsche
  Forschungsgemeinschaft within the Emmy-Noether grant KO 2302/1-1 (CPK)
  and within SFB 450 (MN,RK)  is gratefully acknowledged.
  The Fritz Haber
  Center is supported by the Minerva Gesellschaft f\"{u}r die Forschung
  GmbH M\"{u}nchen, Germany.
\end{acknowledgments}

\appendix

\section{Transformation to obtain $|\Phi^{(j)}  \rangle$ from
  $|\bar{\Phi}_j  \rangle$ } 
\label{app:trafo}

When the solution of the inhomogeneous Schr\"odinger equation is
based on the uniform approximation, a transformation linking
the Chebychev expansion coefficients, $|\bar{\Phi}_j\rangle$, to the
coefficients of the Taylor expansion,  $|\Phi^{(j)}  \rangle$,
cf. Eq.~(\ref{eq:Taylor}), is required. In other words, 
given a vector $[A_0,\ldots A_m]^T$ we want to compute the vector
$[B_0,\ldots B_m]^T$ such that 
\begin{equation}
  \sum_{k=0}^{m} A_{m,k}P_k(x) =  \sum_{j=0}^{m} B_{m,j}\frac{x^j}{j!}\,, 
\label{eq:AnBn}
\end{equation} 
i.e. we identify $|\bar{\Phi}_k  \rangle$ and $|\Phi^{(j)}  \rangle$ to
$A_{m,k}$ and  $B_{m,j}$ respectively. 
Let 
\begin{equation}
  P_k(x) =  \sum_{j=0}^{k} C_{k,j}\frac{x^j}{j!}\,. 
\label{eq:chebypol}
\end{equation} 
Since Chebychev polynomials obey the recursion relation, 
\begin{equation}
  P_{k+1}(x) = 2x P_k(x) -  P_{k-1}(x) \,, 
\label{eq:chebyrecur}
\end{equation} 
we obtain
\begin{equation}
 \sum_{j=0}^{k+1} C_{k+1,j}\frac{x^j}{j!} =  2 \sum_{j=0}^{k}
 C_{k,j}\frac{x^{j+1}}{j!} - \sum_{j=0}^{k-1} C_{k-1,j}\frac{x^j}{j!}
 \,,  
\label{eq:coeffck1}
\end{equation} 
or
\begin{equation}
 \sum_{j=0}^{k+1} C_{k+1,j}\frac{x^j}{j!} =  2 \sum_{j=1}^{k+1}
 C_{k,j-1}\frac{x^j}{(j-1)!} - \sum_{j=0}^{k-1} C_{k-1,j}\frac{x^j}{j!}
 \,.  
\label{eq:coeffck2}
\end{equation} 
Hence, the $C$ coefficients satisfy 
\begin{eqnarray}
 C_{k+1,0} &=&  -C_{k-1,0}\,, \nonumber \\
 C_{k+1,j} &=& 2 j C_{k,j-1} -  C_{k-1,j} \quad,\quad  1\le j \le k-1  \nonumber  \\
 C_{k+1,k} &=& 2k C_{k,k-1} \,, \nonumber  \\
 C_{k+1,k+1} &=& 2(k+1) C_{k,k} \,.
\end{eqnarray}
Based on this result we can compute the $B$ coefficients recursively,
\begin{eqnarray}
 \label{eq:coeffBi}
 B_{i+1,j} &=& B_{i,j} + A_{i+1,i+1}C_{i+1,j}  \quad,\quad 1\le j \le i \nonumber \\
   B_{i+1,i+1} &=& A_{i,i}  C_{i,i} \,,
\end{eqnarray}
for $ 1\le i \le m-1$ with
\begin{equation}
\label{eq:coeffB0}
 B_{0,0} =  A_{0,0}\quad,\quad   B_{1,0} =  A_{1,0}\quad,\quad B_{1,1} =  A_{1,1}\;. 
\end{equation}

\section{Proof of the equivalence of Eqs.~(\ref{eq:formalsol})
  and (\ref{eq:formalsolbis})} 
\label{app:approx}
It is shown by induction that the formal solution, 
Eq.~(\ref{eq:formalsol}), and Eq.~(\ref{eq:formalsolbis}) are equivalent.
%
Writing Eqs.~(\ref{eq:formalsol}) and (\ref{eq:formalsolbis})
for $m = 1$, one obviously obtains in both cases 
the equation of the first order, Eq.~(\ref{eq:firstorder}). 
We assume that  Eqs.~(\ref{eq:formalsol}) and (\ref{eq:formalsolbis})
are equivalent in order $m-1$, i.e,
\begin{widetext}
\begin{equation}
\label{eq:proofm-1}
  |\psi(t)\rangle_{(m-1)} = \sum_{j=0}^{m-2} \frac{t^j}{j!} |\lambda^{(j)} \rangle +
  \Op{F}_{m-1} |\lambda^{(m-1)}\rangle \,
   =  e^{-i\Op{H} t} |\psi_0 \rangle +
  \sum_{j=0}^{m-2}  \Op{F}_{j+1}  |\Phi^{(j)}  \rangle\,,
\end{equation}
let us now prove that they are equivalent in order $m$.
\begin{eqnarray}
\label{eq:proofm}
 |\psi(t)\rangle_{(m)}  &=& \sum_{j=0}^{m-1} \frac{t^j}{j!} |\lambda^{(j)} \rangle +
  \Op{F}_{m} |\lambda^{(m)}\rangle  \nonumber\, \\
 &=&  \sum_{j=0}^{m-2} \frac{t^j}{j!} |\lambda^{(j)} \rangle + 
\frac{t^{m-1}}{(m-1)!} |\lambda^{(m-1)} \rangle +
 \Op{F}_{m} \left( (-i\Op{H})|\lambda^{(m-1)}\rangle +  
          |\Phi^{(m-1)}  \rangle \right) \nonumber \, \\
 &=&  \sum_{j=0}^{m-2} \frac{t^j}{j!} |\lambda^{(j)} \rangle +
        \Op{F}_{m}   |\Phi^{(m-1)}  \rangle +
       \left(\Op{F}_{m}  (-i\Op{H}) +  \openone
         \frac{t^{m-1}}{(m-1)!}\right)|\lambda^{(m-1)}\rangle    
\end{eqnarray}
We continue by showing that
$ \Op{F}_{m-1} = \Op{F}_{m}  (-i\Op{H}) +   \openone \frac{t^{m-1}}{(m-1)!}$,
\begin{eqnarray*}
  \Op{F}_{m}  (-i\Op{H})   +  \openone \frac{t^{m-1}}{(m-1)!} 
  &=&  (-i\Op{H})^{-(m-1)} \left (e^{-i\Op{H} t} -
    \sum_{j=0}^{m-2} \frac{(-i \Op{H} t)^j}{j!} -  
    \frac{(-i\Op{H}t)^{m-1}}{(m-1)!} \right)  +  \openone\frac{t^{m-1}}{(m-1)!}\, \\
  &=&  (-i\Op{H})^{-(m-1)} \left (e^{-i\Op{H} t} -
    \sum_{j=0}^{m-2} \frac{(-i \Op{H} t)^j}{j!}  \right) 
  - (-i\Op{H})^{-(m-1)} \frac{(-i\Op{H}t)^{m-1}}{(m-1)!}  +   \openone
  \frac{t^{m-1}}{(m-1)!}\, \\ 
  &=&   (-i\Op{H})^{-(m-1)} \left (e^{-i\Op{H} t} -
    \sum_{j=0}^{m-2} \frac{(-i \Op{H} t)^j}{j!}  \right)-
 \openone( \frac{t^{m-1}}{(m-1)!}
  - \frac{t^{m-1}}{(m-1)!}) \, \\
  &=&  (-i\Op{H})^{-(m-1)} \left (e^{-i\Op{H} t} -
    \sum_{j=0}^{m-2} \frac{(-i \Op{H} t)^j}{j!}  \right) =
  \Op{F}_{m-1} \,. 
\end{eqnarray*}
Equation (\ref{eq:proofm}) thus becomes
\begin{eqnarray*}
  |\psi(t)\rangle_{(m)} 
  =  \sum_{j=0}^{m-2} \frac{t^j}{j!} |\lambda^{(j)} \rangle +
  \Op{F}_{m-1} |\lambda^{(m-1)}\rangle 
  + \Op{F}_{m}   |\Phi^{(m-1)}  \rangle \,. 
\end{eqnarray*}
Making use of our assumption, Eq.~(\ref{eq:proofm-1}), we obtain
\begin{eqnarray*}
  |\psi(t)\rangle_{(m)} 
  &=&  e^{-i\Op{H} t} |\psi_0 \rangle +
  \sum_{j=0}^{m-2}  \Op{F}_{j+1}  |\Phi^{(j)}  \rangle\
  + \Op{F}_{m}   |\Phi^{(m-1)}  \rangle \\ &=& 
  e^{-i\Op{H} t} |\psi_0 \rangle +
  \sum_{j=0}^{m-1}  \Op{F}_{j+1}  |\Phi^{(j)}  \rangle\,.
\end{eqnarray*}
This concludes the proof.
\end{widetext}

\section{Algorithm of  the symmetrical method}
\label{app:symm_method}

The symmetrical method of Ref.~\cite{JoseMyPRA08} is based on the
formal integration of the inhomogeneous time-dependent Schr\"odinger
equation, Eq.~(\ref{eq:inhSE}),
\begin{equation}
  \label{eq:formalsolsymm}
  |\psi(t)\rangle =  e^{-i\Op{H}t} |\psi(0)\rangle + 
e^{-i\Op{H}t}  \int_0^{t} e^{i\Op{H}\tau} |\bar{\Phi}
(\tau)\rangle d\tau \,. 
\end{equation}
Assuming $ |\bar{\Phi}(\tau)\rangle$ to be constant in $[0,t]$ and taking
its value to be
\begin{equation}
|\bar{\Phi} (\tau)\rangle = \frac{|\Phi (0)\rangle + |\Phi
  (t)\rangle}{2} \qquad \forall \tau \in [0,t] \,,
\label{eq:phiconst}
\end{equation}
the integral in Eq.~(\ref{eq:formalsolsymm}) can easily be computed
and we obtain
\begin{equation}
  \label{eq:formalsolsymmbis}
  |\psi(t)\rangle \approx  e^{-i\Op{H}t} |\psi(0)\rangle + 
  (-i\Op{H})^{-1}\left(e^{-i\Op{H}t} - \openone \right)  |\bar{\Phi} (0)\rangle\,. 
\end{equation}
This solution is formally equivalent to the first order of the
Chebychev propagator, cf. Eq~(\ref{eq:firstorder}). However,
the evaluation of $|\psi(t)\rangle$ proceeds differently in the
symmetrical method and the first order Chebychev propagator. The
latter subjects Eq.~(\ref{eq:formalsolsymmbis}) to a spectral
approximation. It only requires a representation of the Hamiltonian such
that its action on a state vector can be evaluated. A numerically very
efficient representation is based on the Fourier grid where 
evaluation of $\Op{H}|\psi\rangle$ scales as  $O(N\log N)$ with $N$
the number of grid points.  
The symmetrical method diagonalizes $\Op{H}(t)$ at each time step in order
to directly employ Eq.~(\ref{eq:formalsolsymmbis}). Since
diagonalization scales as $O(N^3)$, where $N$ is the dimension of the
Hilbert space, this is feasible only for sufficiently small $N$.

\end{document}